\documentclass[journal]{IEEEtran}
\usepackage{blindtext, graphicx}

\usepackage{balance}  
\usepackage{fixltx2e}
\usepackage{algorithmicx}
\usepackage{algorithm}
\usepackage[noend]{algpseudocode}
\usepackage{url}
\let\labelindent\relax
\usepackage{enumitem}
\usepackage{multirow}
\usepackage{graphicx}
\usepackage{balance}  
\usepackage{url}
\usepackage[acronym,toc,shortcuts]{glossaries}
\usepackage{epsfig}
\usepackage{amssymb}
\usepackage{amsthm}
\usepackage{multirow}
\usepackage{caption}
\usepackage{subcaption}
\usepackage{siunitx}
\usepackage{soul}
\usepackage{algorithm}
\usepackage{algpseudocode}
\usepackage{csquotes}
\usepackage{bbm}
\usepackage{dsfont}
\usepackage{xcolor}
\usepackage{enumitem}
\usepackage{adjustbox}
\usepackage{array}
\usepackage{multirow}
\usepackage{fancyhdr}
\usepackage[T1]{fontenc}
\usepackage{balance}

\usepackage[acronym,toc,shortcuts]{glossaries}

\graphicspath{{figures/}}

\newacronym{3GPP}{3GPP}{The 3rd Generation Partnership Project }
\newacronym{5G}{5G}{Fifth Generation}
\newacronym{6G}{6G}{Sixth Generation}
\newacronym{AAA}{AAA}{Authentication, Authorization and Accounting}
\newacronym{AES}{AES}{Advanced Encryption System}
\newacronym{AI}{AI}{Artificial Intelligence}
\newacronym{AP}{AP}{Access Point}
\newacronym{API}{API}{Application Programming Interface}
\newacronym{APN}{APN}{Access Point Name}
\newacronym{AR}{AR}{Augmented Reality}
\newacronym{BS}{BS}{Base Station}
\newacronym{BER}{BER}{Bit Error Rate}
\newacronym{BSSID}{BSSID}{Basic Service Set Identification}
\newacronym{CAT}{CAT}{Capacity-Aware TOPSIS}
\newacronym{CEP}{CEP}{Complex Event Processing}
\newacronym{CELL-ID}{CELL-ID}{cell identification ID}
\newacronym{CGI}{CGI}{Cell Global Identification}
\newacronym{CLSM}{CLSM}{Closed loop spatial multiplexing}
\newacronym{CQI}{CQI}{Channel Quality Indicator}
\newacronym{CN}{CN}{core network}
\newacronym{CNN}{CNN}{Convolutional Neural Networks}
\newacronym{CL}{CL}{Closed-Loop}
\newacronym{CoMP}{CoMP}{coordinated multi-point}
\newacronym{CS}{CS}{central scheduler}
\newacronym{CSI}{CSI}{Channel Status Information}
\newacronym{CPU}{CPU}{Central Processing Unit}
\newacronym{eNB}{eNB}{evolved Node-B}
\newacronym{DL}{DL}{Downlink}
\newacronym{DES}{DES}{Data Encryption Standard}
\newacronym{DMM}{DMM}{Distributed Mobility Management}
\newacronym{DoS}{DoS}{Denial of Service}
\newacronym{DTLS}{DTLS}{Datagram Transport Layer Security}
\newacronym{EC}{EC}{Edge Computing}
\newacronym{ECA}{ECA}{Event-Condition-Action}
\newacronym{ECC}{ECC}{Elliptic Curve Cryptography}
\newacronym{eNodeB}{eNodeB}{evolved Node-B}
\newacronym{E-RAB}{E-RAB}{E-UTRAN Radio Access Bearer}
\newacronym{ETSI}{ETSI}{European Telecommunications Standards Institute}
\newacronym{FDD}{FDD}{Frequency Division Duplexing }
\newacronym{FEM}{FEM}{Flow Extraction Manager}
\newacronym{GGSN}{GGSN}{Gateway GPRS Support Node}
\newacronym{GPRS}{GPRS}{General packet radio service}
\newacronym{GTP}{GTP}{GPRS Tunneling Protocol}
\newacronym{HetNet}{HetNet}{heterogeneous network}
\newacronym{HSS}{HSS}{Home Subscriber Station}
\newacronym{HTTP}{HTTP}{Hypertext Transfer Protocol}
\newacronym{HTTPS}{HTTPS}{Hypertext Transfer Protocol Secure}
\newacronym{HDFS}{HDFS}{Hadoop Distributed File System}
\newacronym{HiveQL}{HiveQL}{Hive Query language}
\newacronym{HSPA}{HSPA}{High Speed Packet Access}
\newacronym{IBLER}{IBLER}{Initial Block Error Rate}
\newacronym{ICIC}{ICIC}{inter-cell interference coordination}
\newacronym{ICN}{ICN}{information-centric network}
\newacronym{IEEE}{IEEE}{Institute of Electrical and Electronics Engineers}
\newacronym{IETF}{IETF}{Internet Engineering Task Force}
\newacronym{IMSI}{IMSI}{International Mobile Subscriber Identity}
\newacronym{IMEI}{IMEI}{International Mobile Station Equipment Identity}
\newacronym{IMS}{IMS}{IP Multimedia Subsystem}
\newacronym{ICMP}{ICMP}{Internet Control Message Protocol}
\newacronym{IoT}{IoT}{Internet of Things}
\newacronym{IP}{IP}{Internet Protocol}
\newacronym{IPSec}{IPSec}{Internet Protocol Security}
\newacronym{ITU}{ITU}{International Telecommunication Union}
\newacronym{IT}{IT}{Information Technology}
\newacronym{GBR}{GBR}{Guaranteed Bit Rate}
\newacronym{GLUE}{GLUE}{General Language Understanding Evaluation}
\newacronym{JSON}{JSON}{JavaScript Object Notation}
\newacronym{KPI}{KPI}{Key Performance Indicator}
\newacronym{LA}{LA}{Location Area}
\newacronym{LAC}{LAC}{location area code}
\newacronym{LMA}{LMA}{Local Mobility Anchor}
\newacronym{LTE}{LTE}{long term evolution}
\newacronym{MADM}{MADM}{Multiple Attribute Decision Making}
\newacronym{MANO}{MANO}{Management and Orchestration}
\newacronym{MCC}{MCC}{Mobile Country Code}
\newacronym{MEC}{MEC}{Mobile Edge Computing}
\newacronym{MCS}{MCS}{Modulation Coding Scheme}
\newacronym{MCP}{MCP}{Management Control Policy}
\newacronym{MNC}{MNC}{Mobile Network Code}
\newacronym{MIMO}{MIMO}{multiple-input multiple-output}
\newacronym{MAG}{MAG}{Mobile Access Gateway}
\newacronym{MAAR}{MAAR}{Mobility Anchor and Access Router}
\newacronym{ML}{ML}{Machine Learning}
\newacronym{MME}{MME}{Mobility Management Entity}
\newacronym{MN}{MN}{Mobile Node}
\newacronym{MNO}{MNO}{Mobile Network Operator}
\newacronym{MSISDN}{MSISDN}{Mobile Station International Subscriber Directory Number}
\newacronym{NBI}{NBI}{NorthBound Interface}
\newacronym{NE}{NE}{Network Equipment}
\newacronym{NFV}{NFV}{Network Function Virtualization}
\newacronym{NIST}{NIST}{National Institute of Standards and Technology}
\newacronym{NLP}{NLP}{Natural Language Processing}
\newacronym{NLU}{NLU}{Natural Language Understanding}
\newacronym{NOMA}{NOMA}{Non-Orthogonal Multiple Access}
\newacronym{NoSQL}{NoSQL}{Not Only SQL}
\newacronym{NR}{NR}{New Radio}
\newacronym{NS}{NS}{Network Service}
\newacronym{QoS}{QoS}{quality-of-service}
\newacronym{QoE}{QoE}{quality-of-experience}
\newacronym{OAM}{OAM}{Operation, Administration and Management}
\newacronym{ONF}{ONF}{Open Networking Foundation}
\newacronym{ONOS}{ONOS}{Open Network Operating System}
\newacronym{OS}{OS}{operating system}
\newacronym{OL}{OL}{Open-Loop}
\newacronym{PDN}{PDN}{packet data network}
\newacronym{PF}{PF}{Proportional Fair}
\newacronym{P-GW}{P-GW}{packet gateway}
\newacronym{PDP}{PDP}{Packet Data Protocol}
\newacronym{PHY}{PHY}{physical layer}
\newacronym{PKI}{PKI}{Public Key Infrastructure}
\newacronym{PMIPv6}{PMIPv6}{Proxy Mobile IPv6}
\newacronym{PMI}{PMI}{Precoding Matrix Index}
\newacronym{PRB}{PRB}{Physical Resource Block}
\newacronym{PUSCH}{PUSCH}{Physical Uplink Shared Channel}
\newacronym{REST}{REST}{Representational State Transfer}
\newacronym{QAM}{QAM}{Quadrature amplitude modulation}
\newacronym{QCI}{QCI}{QoS Class Identifier}
\newacronym{QRSA}{QRSA}{Quantum Resistant Security Algorithm}
\newacronym{RA}{RA}{Routing Area}
\newacronym{RB}{RB}{Resource Block}
\newacronym{RI}{RI}{Rank Indicator}
\newacronym{RAN}{RAN}{radio access network}
\newacronym{RFC}{RFC}{Request for Comment}
\newacronym{RRC}{RRC}{Radio Resource Control}
\newacronym{RNC}{RNC}{radio network controller}
\newacronym{RNN}{RNN}{Recurrent Neural Networks}
\newacronym{RSSI}{RSSI}{Received Signal Strength Indicator}
\newacronym{RSRP}{RSRP}{Reference Signal Received Power}
\newacronym{OTT}{OTT}{over-the-top}
\newacronym{SA}{SA}{Stand Alone}
\newacronym{SAC}{SAC}{service area code}
\newacronym{SCMA}{SCMA}{Sparse Code Multiple Access}
\newacronym{SLA}{SLA}{Service Level Agreement }
\newacronym{SDN}{SDN}{Software Defined Networking}
\newacronym{SDO}{SDO}{Standards Developing Organization}
\newacronym{SFN}{SFN}{Single Frequency Network}
\newacronym{S-GW}{S-GW}{serving gateway}
\newacronym{SINR}{SINR}{signal-to-interference-plus-noise ratio}
\newacronym{SGSN}{SGSN}{Serving GPRS Support Node}
\newacronym{SCO}{SCO}{Service \& Computation Orchestrator}
\newacronym{SSID}{SSID}{Service Set Identification}
\newacronym{SVD}{SVD}{singular value decomposition}
\newacronym{TCP}{TCP}{transport control protocol}
\newacronym{TDD}{TDD}{Time Division Duplexing}
\newacronym{TLS}{TLS}{Transport Layer Security}
\newacronym{TM}{TM}{transmission mode}
\newacronym{TEID}{TEID}{tunnel endpoint identifier}
\newacronym{UDN}{UDN}{Ultra Dense Network}
\newacronym{UMTS}{UMTS}{Universal Mobile Telecommunications Service} 
\newacronym{UE}{UE}{user equipment}
\newacronym{UL}{UL}{Uplink}
\newacronym{UDP}{UDP}{User Datagram Protocol}
\newacronym{V2X}{V2X}{Vehicle-to-everything}
\newacronym{VNF}{VNF}{Virtual Network Function}
\newacronym{WiFi}{WiFi}{Wireless Fidelity}
\newacronym{WLAN}{WLAN}{Wireless Local Area Network}

\begin{document}
%
\title{Post-Quantum Era in V2X Security:  Convergence of Orchestration and Parallel Computation}

\author{Engin~Zeydan$^{\diamond}$, Yekta Turk$^{\ast}$, Berkin Aksoy$^{\ast}$ and Yaman Yagiz Tasbag$^{\ast}$  \\
 $^{\diamond} $Centre Tecnològic de Telecomunicacions de Catalunya (CTTC), Castelldefels, Barcelona, Spain, 08860. \\
 $^{\ast} $Aselsan Corp., Istanbul, Turkey, 34396. \\
\protect engin.zeydan@cttc.cat, [yektaturk, berkinaksoy, yytasbag]@aselsan.com.tr}

\maketitle

\begin{abstract}

Along with the potential emergence of quantum computing, safety and security of new and complex communication services such as automated driving need to be redefined in the post-quantum era. To ensure reliable, continuous and secure operation of these scenarios, quantum resistant security algorithms (QRSAs) that enable secure connectivity must be integrated into the network management and orchestration systems of mobile networks. This paper explores a roadmap study of post-quantum era convergence with cellular connectivity using the Service \& Computation Orchestrator (SCO) framework for enhanced data security in radio access and backhaul transmission with a particular focus on Vehicle-to-Everything (V2X) services.  Using NTRU as a QSRA, we have shown that the parallelization performance of Toom-Cook and Karatsuba computation methods can vary based on different CPU load conditions through extensive simulations and that the SCO framework can facilitate the selection of the most efficient computation for a given QRSA. Finally, we discuss the  evaluation results, identify the current standardization efforts, and possible directions for the coexistence of post-quantum  and mobile network connectivity through a SCO framework  that leverages parallel computing.

\end{abstract}


\IEEEpeerreviewmaketitle

\section{Introduction}

Next generation cellular communication systems are expected to further drive the diversity of applications provided by 5G systems \cite{yazar6g}. They will provide a highly flexible platform that can can deliver features such as integrated sensing, flexible multi-band usage, smart, green and secure communications.  In next-generation cellular communication,  \ac{AI}, integrated sensing and communication are expected to be key enablers. \ac{V2X} or autonomous driving has already attracted much attention in recent years and is expected to be used in next-generation communication systems. For example, convergence of \ac{MEC} and blockchain technologies for a novel robust, resilient, and reliable architecture for V2X communications is proposed in \cite{khan2021robust}. With the recent industrial achievements from Google's Waymo, Baidu's Apollo and  Tesla's Autopilot technologies \cite{Cusumano}, full self-driving capabilities  are possible in the future. However, the underlying secure communication technologies that support them  are still in their infancy.

In the field of security, quantum computers are slowly being developed by many high profile \ac{IT} and cloud companies (Google, IBM, Honeywell and Microsoft). In the long terms, quantum computers are expected to achieve quantum superiority and recent industrial work by Google has shown that these claims are feasible \cite{arute2019quantum}.  Devices and computers with quantum capacity have the ability to respond very quickly to complex problems by performing many operations simultaneously.  Quantum-capable devices can also be used to attack security protocols established between network devices (such as \ac{TLS}, \ac{DTLS}, \ac{IPSec}, etc.), break encryption schemes used in security protocols, and expose secure communications. In this case,  algorithms resistant to quantum flows in security protocols operated between mobile network devices must be used. Thus, secure communication can be established in the mobile network. \ac{NIST} is currently working on this issue and has organized a competition to select the best algorithm that is resistant to quantum attacks. Algorithms such as NTRU, Saber and Kyber have already made it to the final round of the \ac{NIST} competition \cite{NIST_comp}. After the \ac{NIST} competition is over, one algorithm will be selected as the standard \ac{QRSA} and  implemented in all devices as a result (such as \ac{AES}/\ac{DES}, which are the result of the previous \ac{NIST} competition and are widely used today).

Nowadays, current encryption algorithms can be implemented in software and hardware of mobile devices to ensure the operation of the security protocol. However, a major problem arises when implementing \glspl{QRSA}. This problem is to decide how to perform the computational processes within the algorithm. In the study of existing \glspl{QRSA}  such as NTRU, Saber, Kyber, the load on the \ac{CPU} cores of the devices for computing mathematical operations with very high-order polynomials is very high. The computational processes for existing encryption algorithms such as \ac{AES}, \ac{ECC} etc. are already in use and  the existing hardware capabilities of the devices can support these types of operations. However for \glspl{QRSA}, we are dealing with very high-degree polynomials. Therefore, the computational resources required for \glspl{QRSA} on the devices are higher than the resources required for the encryption algorithms already in use. For this reason, it is very difficult to perform computations for a \ac{QRSA}  based on high-order polynomials with the currently available software and hardware. To overcome this problem, an efficient computation and orchestration framework is needed that is aware of the entire topological view of the mobile and \ac{V2X} network and can perform different computation methods for each \ac{QRSA}. To this end, the following contributions are made in this paper:

\begin{itemize}
    \item We discuss key concepts and present the integration of post-quantum  and mobile network connectivity through the \ac{SCO} framework  with potential use cases. (Section \ref{Background} and \ref{frameworks})
    \item We evaluate how post-quantum methods can be integrated into mobile networks via simulation analysis and discuss the evaluation results.(Section \ref{experiments})
    \item From a standardization perspective, we describe current \glspl{SDO} efforts and  their potential contributions to network management and orchestration and post-quantum technologies. (Section \ref{standardization})
\end{itemize}

\section{Background \& Motivation}
\label{Background}

\subsection{The Approach to Facilitate the Computation}
\label{Approach_Computation}

A \ac{QRSA} is actually based on multiplication of higher order polynomials. This can also be observed when algorithms are studied for \ac{NIST} competition for selecting the next standard \ac{QRSA} such as NTRU, Saber, Kyber, etc.. Multiplication of higher order polynomials consists of dozens of fundamental mathematical arithmetic operations such as addition and multiplication. Among these arithmetic operations, multiplication operations cause problems and implementation bottlenecks \cite{toomy}.  Multiplication processes that take place within the \ac{QRSA}, consume many \ac{CPU} resources and occupy significant \ac{CPU} caches such as L1 and L2 caches.

On the other hand, time complexity is another problem that varies depending on the platform implemented and space complexity is critical for execution in hardware \cite{clifford}. In the real world of device system architectures, we need to target time complexity, which means that shorter execution time is crucial for a mobile node. To overcome this, alternative multiplication operations can be used in software/hardware. For example, in the literature, there are various multiplication methods  such as Karatsuba \cite{karatsuba}, Toom-Cook \cite{toom-cook}, Schönhage–Strassen \cite{schonage}, etc., which can perform the multiplication operations efficiently and with lower time complexity. As the degree of polynomials increases, the computation of  multiplication of two polynomials becomes much more time consuming.  The common feature of these different multiplication methods is that they focus on reducing the number of fundamental arithmetic multiplication operations when multiplying polynomials. This reduces the time required to compute the result of polynomial multiplication.

\begin{figure}[tb!]
\centering
\begin{subfigure}{\linewidth}
  \centering
\includegraphics[width=.9\linewidth]{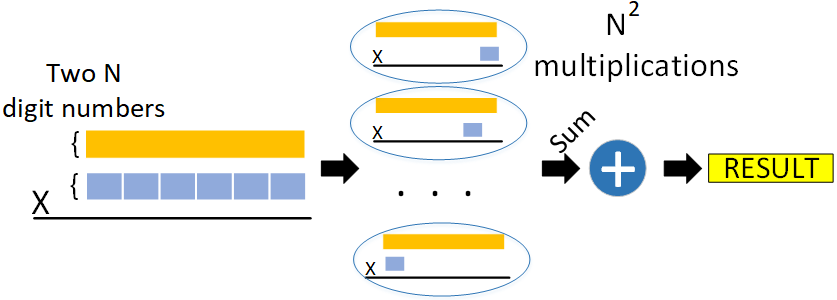}
  \caption{}
  \label{Fig2a}
\end{subfigure}
\begin{subfigure}{\linewidth}
  \centering
\includegraphics[width=\linewidth]{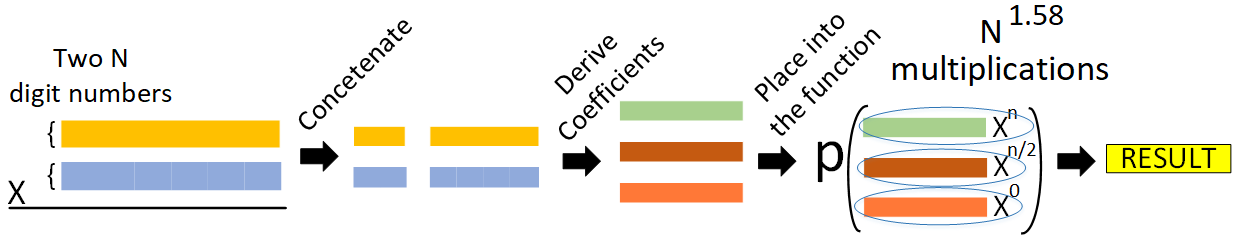}
  \caption{}
  \label{Fig2b}
\end{subfigure}
\begin{subfigure}{\linewidth}
  \centering
\includegraphics[width=\linewidth]{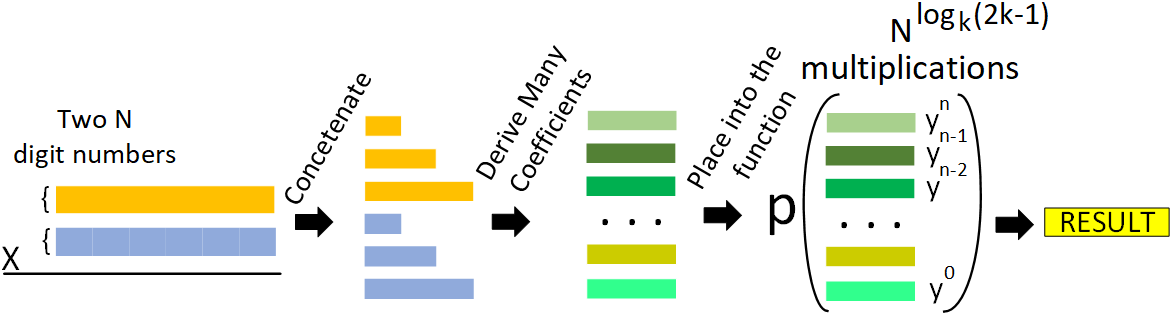}
  \caption{}
  \label{Fig2c}
\end{subfigure}
\caption{Illustration for multiplication of two N-digit numbers by different methods and the number of fundamental multiplication operations by (a) Classical schoolbook method (b) Karatsuba's method (c) Toom-Cook k-way  method.}
\label{Fig2}
\end{figure}

\subsection{Different Methods for Implementing Multiplication}
\label{Matrice}

As an example, consider the product of two numbers according to the classical schoolbook method shown in Fig. \ref{Fig2}. This figure is intended to show that some multiplication methods can be computed in parallel on \ac{CPU} cores (e.g., the Toom-Cook method) while some others (e.g., the Karatsuba method) cannot be computed in a distributed fashion and work well only  on a single \ac{CPU} core. The higher degree of polynomials to be multiplied, the higher the number of fundamental arithmetic multiplications. This is illustrated in Fig. \ref{Fig2a}. When two \textit{N}-digit numbers are multiplied over finite field, a total of $N^2$ multiplication operations and $2N-3$ addition operations are performed using the classical schoolbook multiplication method. The fundamental arithmetic multiplications mentioned above are those performed inside the ellipses in Fig. \ref{Fig2}. Another method for multiplication is that of the Karatsuba shown in Fig. \ref{Fig2b}. The approach of Karatsuba's method is based on dividing numbers into smaller parts. In Karatsuba's method, polynomials are first divided into two smaller parts, which are represented as polynomials. Then,  coefficients are derived from these polynomials. There is also another predefined polynomial \textit{p} and the derived coefficients are then substituted in this polynomial. Thus, the Karatsuba method for multiplications of two-digit numbers requires three-digit multiplications instead of four as in the classical schoolbook case. In general,  the  number  of  fundamental  operations of the Karatsuba method  reduces  to $N^{1.58}$ polynomial multiplications  compared to classical schoolbook multiplication, which has $N^2$.

Toom-Cook's method, which is shown in  Fig. \ref{Fig2c}, is the generalization method of Karatsuba. In Toom-Cook, \textit{k} indicates the number of concatenated parts of the original number. As \textit{k} increases, polynomials are divided into smaller parts (the size of the polynomial decreases). In this way, multiplication results can be achieved in fewer recursion steps (lower time complexity), but the overall space complexity increases. In Toom-Cook k-way method, $N^{\log _{k} (2k-1)}$ multiplications are required. If $k=2$, it becomes the Karatsuba method, which divides the \textit{N}-digit numbers into two parts. Thus, Karatsuba method can also  be considered as a Toom-Cook two-way method.

\subsection{QRSA Implementation Target \& Parallel Computing}
\label{Implementation}

Each multiplication method  has its own advantages and disadvantages. One disadvantage of Toom-Cook is that the multiplication results are obtained with high space complexity, but it has fewer recursion steps than  Karatsuba's method.  Karatsuba's method has lower space complexity but uses a lot of digits as it reduces the degrees of the numbers to be multiplied by half. If a mobile node has many \ac{CPU} cores and most of them are underutilized, then the computation process for the \ac{QRSA} can be completed by using the Toom-Cook's method, which has fewer recursion steps with many operations. In this case, the operations can be solved in a short time with parallel computing by using all or most of the available \ac{CPU} cores. Each of the multiplication operations can be performed by different \glspl{CPU}. 

Suppose additionally that we have a mobile node with a single \ac{CPU} core. In this case, we can reach the result of multiplication faster using Karatsuba's method, which outputs results in many recursion steps with fewer operations. In Karatsuba's method, parallel computing has less impact on the  performance than in Toom-Cook method due to the smaller number of operations per step and the dependence of the step result. Note that the method of multiplication in \ac{CPU} cores in the mobile node may vary depending on the availability of resources in \ac{CPU} cores. Thus, if the most appropriate method can be selected, then the encryption/decryption process introduced by \ac{QRSA} can be computed quickly.

\subsection{Selecting the Computation Method for any QRSA}
\label{problem}

The considered problem is to select the computation method for a \ac{QRSA}, executed inside the mobile nodes and on applications such as \ac{V2X}, drones, etc. connected to the \ac{MEC}. Note that when selecting the most appropriate computation method, all relevant parameters such as the current/future utilization status of the  \ac{CPU}/memory of the device, the status of the network topology, etc. must be taken into account. An important point is that the computations performed for \glspl{QRSA} impose an additional load on the \glspl{NE}. The \ac{MEC} servers in the mobile network must perform multiple arithmetic operations simultaneously. In addition, providing services to air-connected \glspl{UE}/autonomous vehicles that jointly perform  the \ac{QRSA} computations may overwhelm the computational resources of \glspl{NE}. 

In this study, a \ac{SCO} framework is used for efficient operation of systems resistant to post-quantum attacks in next-generation mobile networks.  Along with the system used, \glspl{QRSA} security protocols in mobile networks  can reduce the computational burden by using one of the most relevant computation methods (Toom-Cook or Karatsuba). Since the system used has the potential to reduce the computational load on the device, it also enables the mobile nodes to provide more efficient services within the network through lower latency and more secure communication services. The \ac{SCO} framework obtains information about the \glspl{NE} capabilities by interacting with other systems to organize the computational operations. The most appropriate computational method for executing a \ac{QRSA} is communicated to the \glspl{NE} by monitoring the current load on the \ac{NE}.

\begin{figure*}[htp!]
\centering
\includegraphics[width=\linewidth]{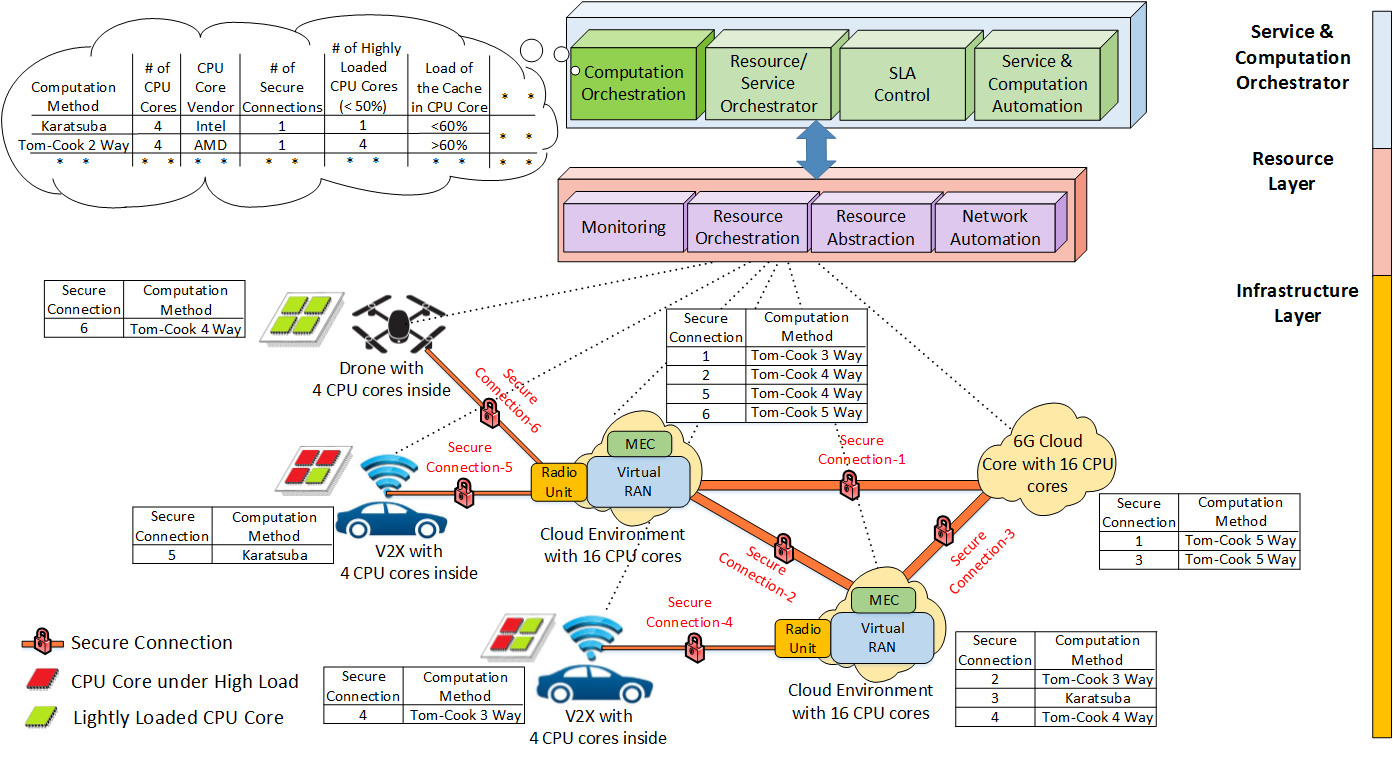}
\caption{An illustration of how a SCO-based framework and post-quantum computing can be architecturally integrated into a mobile network infrastructure.} 
\label{proposed_architecture}
\end{figure*}

\section{Secure Service Computation \& Orchestration in Post-Quantum Era}
\label{frameworks}

The main goal of the \ac{SCO} framework is to guarantee the most efficient computation for the \ac{QRSA}. The perspective is purely from the point of view of computing and acceleration of the computation process for each \ac{QRSA}. As  mentioned  earlier, all \glspl{QRSA} in the NIST competition (NTRU, Saber, Kyber) are based on multiplication of very high order polynomials. NIST’s competition will result in only one \ac{QRSA} to be standardized. Note that if NTRU runs on one NE, Saber cannot run on the other communicating \ac{NE}. This is similar to the case when \ac{AES} is running in a \ac{NE}, \ac{DES} cannot run in a communicating  \ac{NE} while establishing secure communication. For this reason, if NTRU is selected as the winner of the \ac{NIST} competition, it must run on both communicating \glspl{NE}. 

This is where \ac{SCO} comes into play for selecting the appropriate computation method within \ac{QRSA}.  Regardless of which algorithm is chosen by \ac{NIST}, this does not change the requirements for a \ac{SCO}, as all  \ac{NIST} finalist \glspl{QRSA} are based on higher order polynomial multiplications.  Note that the \ac{SCO} framework does not aim to notify both sides in a secure communication to implement the same \ac{QRSA}. In this regard, the \ac{SCO} framework does not select the \ac{QRSA} to be used.  Rather, the \ac{SCO} helps to select the most effective computation method for the implemented \ac{QRSA}. Note that this decision is dynamically made by \ac{SCO} and cannot be easily extended by a new resource scheduling algorithm due to the large data dimension and the dynamics of the environment.  For this reason, \ac{AI}/\ac{ML} techniques or Deep Policy Enforcements (which are basically rule-based) can be used for this purpose.

\subsection{High-Level View of the System Design}
\label{hlv}

Fig. \ref{proposed_architecture} shows the high level view of the system architecture. It consists of three architectural blocks. At the top is \ac{SCO}, which provides smart \ac{NS}, compute, and resource orchestration capabilities to instantiate secure \glspl{NS}. It is also responsible for lifecycle management and control of services and optimizes  resources in mobile  infrastructure and cloud. In the middle is  \textit{resource layer}, which  manages the underlying mobile infrastructure (\ac{RAN}, \ac{EC}, transport and core networks) as a whole, orchestrates the resources, instantiates \glspl{VNF}, interconnects the transport resources together and thanks to the capabilities brought by network controllers and the underlying \ac{NFV} resources, provides a common abstraction view of the managed resources to the \ac{SCO} via the IFA005 interface\footnote{ETSI GS NFV-IFA, \url{https://www.etsi.org/deliver/etsi_gs/NFV-IFA/001_099/005/02.01.01_60/gs_NFV-IFA005v020101p.pdf}}. \ac{SCO} can communicate with the underlying resource layer to obtain the relevant information about the \ac{NE} capabilities and the current computational load statistics of the \ac{NE} resources, such as \ac{CPU} load states, cache status of the \ac{CPU} core, network I/O count, etc. At the lower layer is the \textit{infrastructure layer} with all physical and virtual compute, storage and network resources. It can support reprogrammable mechanisms and algorithms to support enhanced security features.

Changes in network topology and handover updates of many moving vehicles create load on each \ac{NE}. For this reason, the \ac{SCO} stores a table and rules with certain thresholds. In accordance with these rules, the appropriate computation method is chosen using the given rule sets. Fig. \ref{proposed_architecture} shows an example of this structure. \ac{SCO} will forward the messages to the resource layer and the resource layer to the infrastructure devices through a secure communication channel.  \ac{SCO} is also aware of past load conditions, which can also help make more accurate predictions for future load conditions using either rule-based or \ac{AI}/\ac{ML} models.  When \ac{QRSA} is run in the \ac{NE} (note that all computational methods discussed above must be available in the \ac{NE} software), the \ac{SCO} layer  provides the instructions for selecting the most appropriate computational method for the \ac{QRSA} under consideration.

\subsection{A Case Study}
\label{casesudy}

We consider a case where  moving vehicles use a continuous \ac{V2X} service. Each vehicle dynamically uses a different mobile network \glspl{MEC} along its route, so the instantaneous load of \glspl{MEC} varies.  Each vehicle can run multiple services (e.g., critical services, high priority services, and low priority services) on multiple cores simultaneously. For example, critical services may run object recognition, accident prevention and remote control applications, high priority services may use \ac{AR} navigation and information service, low priority services can use caching databases, multimedia or cloud gaming \cite{dalgkitsis2020data}.  For self-driving capability, critical services should at least be available, run without interruptions (i.e., with high \ac{QoS} support, low latency), and provide basic functionalities (e.g., real-time communication, remote monitoring, proactive driving support) for safety and security reasons.  

On the other hand, there are also major threats to the security and privacy of vehicular networks. For example, illegal monitoring of message transmissions in vehicular networks can compromise message confidentiality, broadcast tempering attacks can disrupt message integrity or  \ac{DoS} attacks can flood vehicular control channels with huge amounts of messages \cite{ghosal2020security}. In the post-quantum era, from a  mobile backhaul perspective, user plane communication between the \glspl{BS} (where the V2X service is received by users) and the core network  is encrypted and can be compromised. From a radio access perspective, the security of V2X devices connected to \glspl{BS} is also threatened, as the confidentiality and integrity algorithms used in the radio protocol layers are vulnerable to quantum attacks.

When an autonomous vehicle with \ac{V2X} service continuously handovers to a new \glspl{MEC} on the highway,  the vehicle must receive service from the nearest \ac{MEC} serving that specific new \ac{MEC} due to critical low latency service requirements. In this case, the vehicle must be connected to the new  \ac{MEC} very quickly and securely. However, since the vehicle is in motion, it will have multiple connections to different \glspl{MEC} as it travels down the road, which can cause significant delays. In this case, the computations within the \ac{QRSA} need to be performed repeatedly and  fast re-establishment to each new \ac{MEC} connection is very important for autonomous driving vehicles. From another perspective, multiple vehicles also need to be securely connected to the new \glspl{MEC} within the mobile network. In this case, \ac{QRSA} computations need to be performed dynamically on the \ac{MEC} side. This increases the load on \ac{CPU}, which overloads the \ac{MEC} and can cause significant delays.

In the architecture of Fig. \ref{proposed_architecture}, the \ac{SCO}, which can monitor the overall network status, is able to choose the most appropriate computation methods for a given \ac{QRSA} for vehicles moving towards \ac{MEC}. Later, this decision is propagated to the underlying layers and to the \ac{MEC} and the vehicle. This ultimately leads to shorter execution times and, as a result, fewer delays in handovers while maintaining a high level of vehicle security.

\section{Experimental Results \& Discussions}
\label{experiments}

We used Kubernetes as our candidate \ac{SCO} and utilized its \ac{CPU}  \ac{MCP} feature\footnote{Kubernetes v1.12, Control CPU Management Policies on the Node, \url{https://kubernetes.io/docs/tasks/administer-cluster/cpu-management-policies/}} in the test environment. The nodes are separate workstations with 48 \ac{CPU} cores and 64 GB memory, but the tests are run with only five of the \ac{CPU} cores. The L1 cache of the nodes is 768  kiB, the L2 cache is 24 miB and the L3 cache is 33 miB. The \ac{QRSA} computation applications are implemented in Docker containers in the nodes. Each computation method is implemented in a different container. The load status of the \ac{CPU} cores is regularly monitored by Kubernetes and the reservation of the five \ac{CPU} cores is also done by the \ac{MCP}.

Since Karatsuba's method works better without parallelization, we ran it with only one \ac{CPU} core. We ran the Toom-Cook's method both on one \ac{CPU} core and with parallel programming \cite{DALCIN20051108} across all five \ac{CPU} cores. Toom-Cook running on one CPU core is given for benchmarking purposes as the method running on a Single \ac{CPU} core. However, Karatsuba's method is a better solution in single core and  the advantage of Toom-Cook is parallel processing. 

\subsection{Simulation Results}
\label{results}

\begin{figure} [tb]
\centering
\begin{subfigure}{\linewidth}
  \centering
\includegraphics[width=\linewidth]{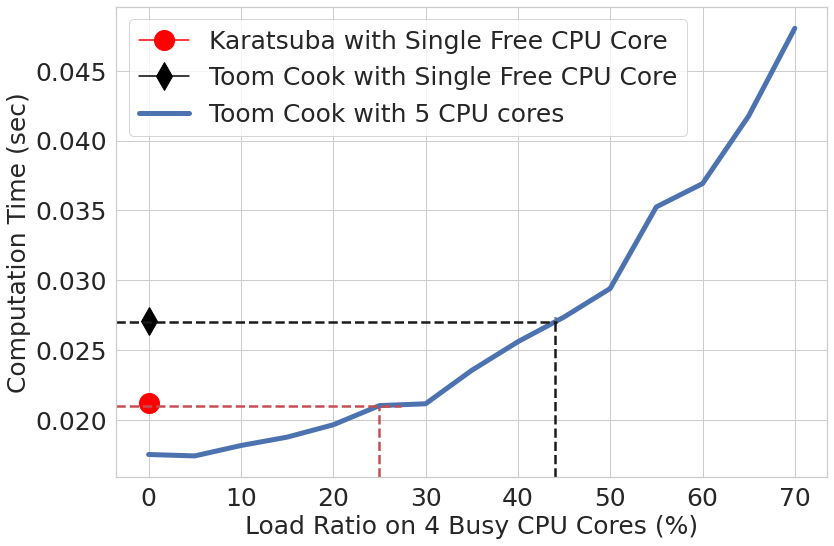}
  \caption{}
  \label{Figsima}\vspace*{4mm}
\end{subfigure}
\begin{subfigure}{\linewidth}
  \centering
\includegraphics[width=\linewidth]{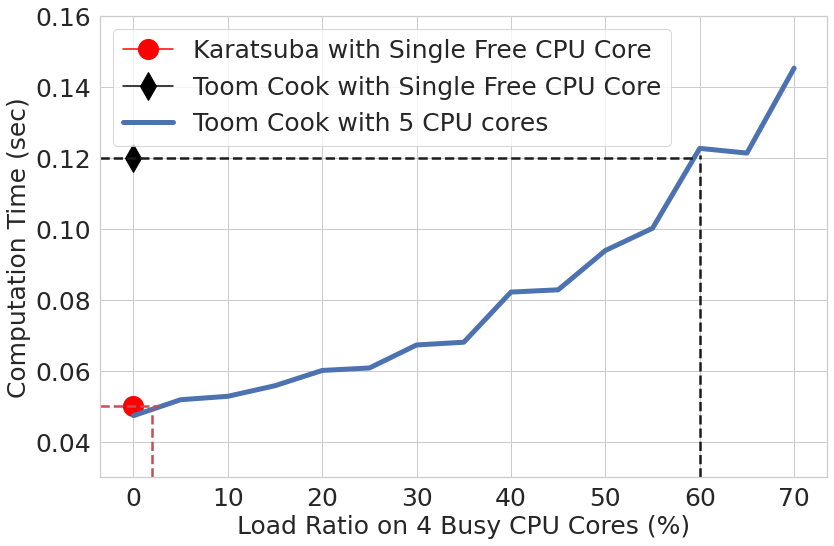}
  \caption{}
  \label{Figsimb}
\end{subfigure}
\caption{Performance comparisons of computational methods under increasing load (load is applied to 4 CPU cores, 1 CPU core is always set with 0\% load) (a) Multiplication of two polynomials with degree 512 (b) Multiplication of two polynomials  with degree 812.}
\label{Figure_simulation}
\end{figure}

Since one \ac{MEC} can perform multiple operations simultaneously in a mobile network, we gradually placed more and more load on three of the \ac{CPU} cores in our simulations. However, we always kept one \ac{CPU} as free (i.e., with 0\% load). Karatsuba's method was always run on this free CPU core. All computation time results are an average of ten Monte Carlo simulations.

Fig. \ref{Figsima} shows the products of two polynomials of degree 512 by different methods. The running \glspl{QRSA} in the NIST's competition generally use polynomials whose degree varies between 256 and 512. For this reason, we studied the product of two polynomials of degree 512 using different multiplication  methods. The results in Fig. \ref{Figsima} show that Karatsuba and Toom-Cook with Single CPU cores yield 21.2 and 27.1 msec computation time, respectively. Moreover, Toom-Cook with five CPU cores performs better than Karatsuba and Toom-Cook with Single CPU cores in terms of computation time up to a load ratio of 25\% on \ac{CPU} cores. After increasing the load from 25\%, the performance of Toom-Cook with five CPU cores starts to decrease gradually compared to Karatsuba. 

Moreover, after 45\% load, the performance of Toom-Cook with five CPU cores becomes even worse than that of Toom-Cook with Single CPU core. The reason is that the resources of the other four \glspl{CPU} are no longer fully reserved for running Toom-Cook. As the load increased, parallelized \ac{CPU} cores could not efficiently handle the multiplication operations involved in each step of Toom-Cook. This is because the computations of arithmetic operations are also heavily loaded on \ac{CPU} cores and the cost of time spent on each step of Toom-Cook increases. Among the \glspl{QRSA} that made it to the 3rd round (finalists) of the \ac{NIST}'s competition, NTRU can achieve  \ac{NIST} security level-5. At level-5, NTRU  has the highest order polynomial multiplication among all \glspl{QRSA} that made it to the \ac{NIST} competition. For this reason, Fig. \ref{Figsimb} also shows the product of two polynomials of degree 821. The results in Fig. \ref{Figsimb} show that Karatsuba and Toom-Cook with Single CPU cores yield 50.2 msec  and 120 msec computation time, respectively. Moreover, in this higher order matrix multiplication case, Karatsuba performs better than Toom-Cook with  five CPU cores in terms of computation time after 5\% load ratio on \ac{CPU} cores and Toom-Cook with Single CPU core performs better than Toom-Cook with five CPU cores  after 60\% \ac{CPU} load.

\subsection{Discussions \& Evaluations of Results}
\label{discussion}

New asymmetric cryptosystems such as Lattice-based, code-based, supersingular elliptic-curve, hash-based and multivariate cryptography are among the leading cryptosystems that are resistant to post-quantum attacks. However, the building blocks of \glspl{QRSA} include matrix-vector products containing high-order polynomials which requires high computational overhead compared to classical computers.  For this reason, it is necessary to design optimal computational techniques.

In Karatsuba's method, the number of recursion steps is high due to polynomial divisions by two. Moreover, the results of these recursion steps are interrelated. For this reason, we cannot take full advantage of the parallel computation. Since each step must wait for the others to complete, the impact of an increased number of \glspl{CPU} on the performance of the parallel computation method is reduced. However, in the Toom-Cook method, the number of recursion steps is small (pros) due to the division of polynomials by $k$ where $k \in  \mathbb{N}_{\geq3}$. Therefore, we can use more CPU cores in each recursion step. For example, if the Toom-Cook $k$-way is used for \ac{QRSA} computation, we need to multiply two polynomials of degree $d$  by dividing them by k (in k recursion steps) and it takes $\log_k(d)$ recursion steps to complete the multiplication. In Karatsuba, on the other hand, it takes $\log_2(d)$. In this way, we can complete Toom-Cook with a smaller number of  steps and make better use of parallel computation.

Since the cost of arithmetic operations required to implement post-quantum algorithms is higher than  the cost of  implementing the algorithms prior to quantum computers, the choice of the optimal multiplication method for each hardware becomes more important. As can be concluded from the above results, the Toom-Cook method, which is more suitable for parallel processing and achieves the multiplication result in a very short time when run in parallel, runs slower than Karatsuba's method when the load ratio on the \ac{CPU} cores increases. A \ac{V2X} device running with five \ac{CPU} cores prefers to compute directly with Toom-Cook, regardless of its load status, which can be misleading. When operating under high load or considering the potentially high load that may occur in the future, switching to Karatsuba's method gives the \ac{V2X} device a significant time advantage. The advantages of the methods over the others differ in different polynomial degrees. The advantages of calculations of different polynomial degrees and the most accurate switching decision time should be known to \ac{SCO}.

\section{Standardization Roadmap} 
\label{standardization}

There have been several works around the world aimed at defining  both network service management/orchestration  and post-quantum technologies. However, they have been worked on separately. \ac{ETSI} \ac{NFV}-\ac{MANO} is one of the main responsible standardization bodies for network service management and orchestration \cite{etsi2014}. On the other hand, quantum  communication has been researched for decades \cite{bassoli2021quantum}. Post-quantum computing technologies conceptually fall into different areas of development separate from the field of  mobile networks \& services domain. 

After the \ac{NIST} competition mentioned in the introduction is completed, we will have a selected and standardized \ac{QRSA}. Therefore, the development of mobile networks and the implementation of post-quantum algorithms require cross-domain collaboration in the industry and extended functional support from the \glspl{SDO} for a reliable \ac{SCO} framework. For this reason,  once \ac{NIST} standardization is completed, \ac{3GPP} and \ac{ETSI} can contribute to the development of post-quantum security integration for mobile network architecture. \ac{IETF} can identify the protocol and \ac{V2X} implementation aspects. Finally, \ac{ITU} can develop guidelines for \glspl{MNO} for the convergence of  post-quantum secured \ac{V2X} services.

For \ac{SCO}-enabled networks to be integrated with future quantum technologies, various computational methods for \glspl{QRSA} (e.g., Karatsba, Toom-cook) should be included in the \ac{SCO} service catalog. The integration points and interfaces also need to be identified, and the corresponding abstractions should be consistent with the current trends in networking, e.g., considering the layered stacks of \ac{ETSI} \ac{MANO} architecture and their modular developments.

\section{Conclusions}
\label{conclusions}

Quantum computing will transform next-generation cellular communications, especially in cryptography, as communication and information flow occur over secure channels. In this paper, we propose a roadmap study for post-quantum era convergence with secure  mobile connectivity via a \ac{SCO} framework, focusing on \ac{V2X} services. Selecting the most appropriate and fastest computation methods of \glspl{QRSA} is an important application area, especially in \ac{V2X} scenarios when the computation of services leading to resource-intensive consumption can be done in parallel.  Through a simulation study, we have shown that it is beneficial to choose the most efficient computation method for \ac{QRSA} when it is triggered by a \ac{SCO}  in terms of computation time and parallelization performance.  Finally, we conclude the paper with  a discussion of the  evaluation  results and  current standardization efforts, and highlight possible directions in the post-quantum era for secure mobile \ac{V2X} network connectivity  from a high-level system design perspective.

\section{Acknowledgment}
This work has been partially funded  by  Generalitat de Catalunya grant 2017 SGR 1195 and the national program on equipment and scientific and technical infrastructure, EQC2018-005257-P under the European Regional Development Fund (FEDER).

\ifCLASSOPTIONcaptionsoff
  \newpage
\fi

\bibliographystyle{IEEEtran}
\bibliography{refs}
%


\begin{IEEEbiography}[{\includegraphics[width=1in,height=1.3in,clip,keepaspectratio]{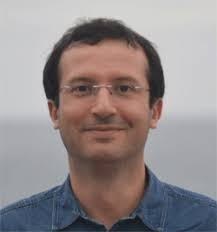}}]{Engin Zeydan} received his Ph.D. degree in Electrical Engineering from Stevens Institute of Technology, Hoboken, NJ, USA, in 2011. He is currently a Senior Researcher at  Centre Tecnologic de Telecomunicacions de Catalunya (CTTC). His research interests are in the areas of telecommunications and data engineering.
\end{IEEEbiography}

\begin{IEEEbiography}[{\includegraphics[width=1in,height=1.3in,clip,keepaspectratio]{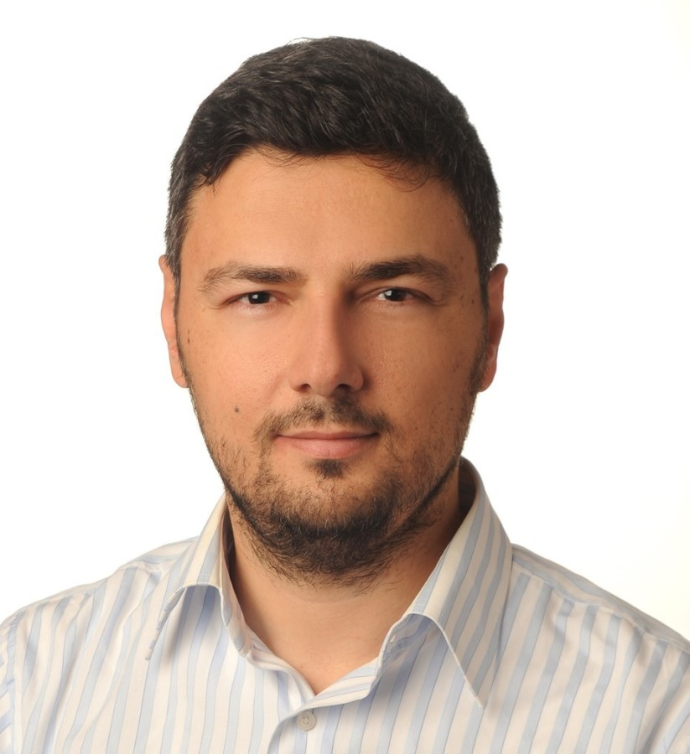}}]{Yekta Turk} received his Ph.D. degree in Computer Engineering from Maltepe University, Istanbul, Turkey, in 2018. He is a Lead Systems Engineer in  Aselsan Corp. His research interests are in the areas of mobile radio telecommunications and computer networks.
\end{IEEEbiography}

\begin{IEEEbiography}[{\includegraphics[width=0.8in,height=1.1in,clip,keepaspectratio]{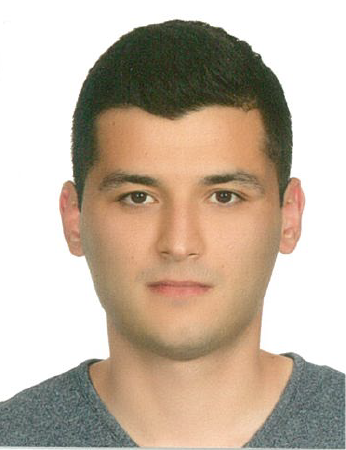}}]{Berkin Aksoy} received his B.Sc degree in Electrical \& Electronics Engineering from Middle East Technical University, Ankara, Turkey, in 2014. He is currently pursuing his M.Sc degree in Institute of Applied Mathematics from the same university. He is a Systems Engineer in  Aselsan Corp. 
\end{IEEEbiography}


\begin{IEEEbiography}[{\includegraphics[width=0.9in,height=1.2in,clip,keepaspectratio]{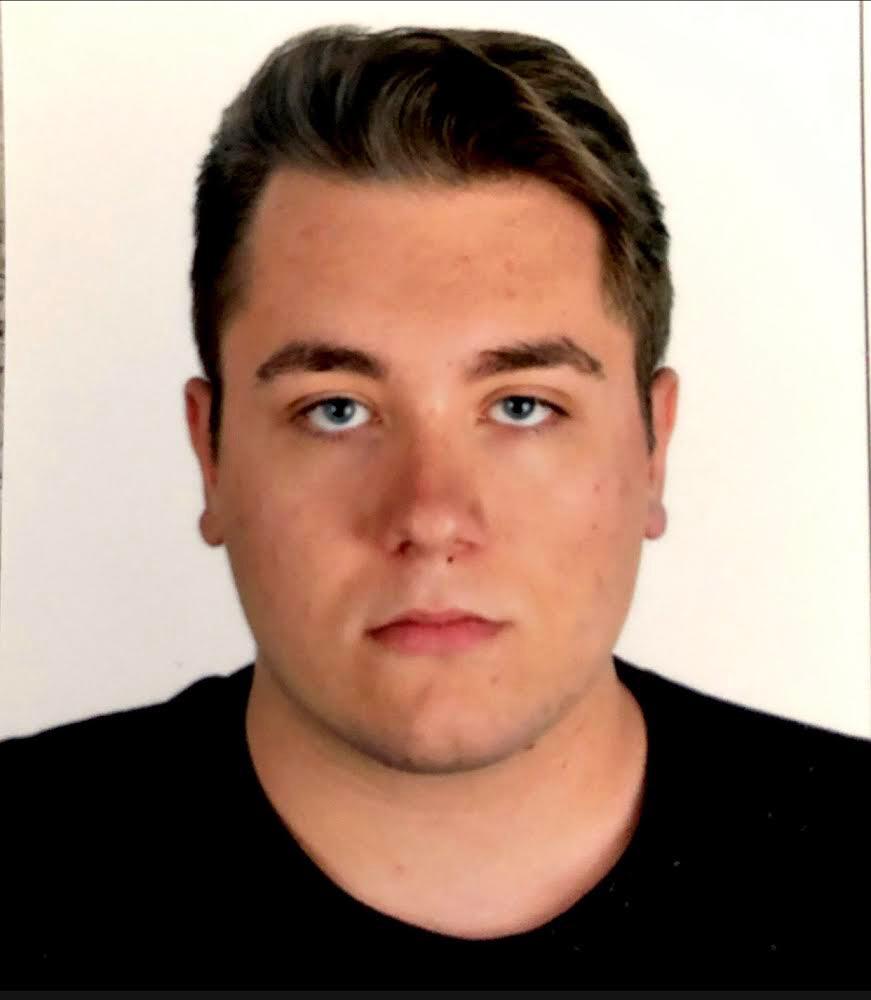}}]{Yaman Yagiz Tasbag} received his B.Sc degree in Computer Engineering from Bilkent University, Ankara, Turkey, in 2020. He is currently pursuing his M.Sc degree from the same university. He is a Systems Engineer in Aselsan Corp. 
\end{IEEEbiography}


\end{document}